\documentstyle[12pt,epsfig]{article}

\newcommand{\beq}{\begin{equation}}
\newcommand{\eeq}{\end{equation}}

\begin{document}

\begin{titlepage}

\begin{center}
\vskip 0.75cm
{\Huge {\bf Strangeness enhancement}}\\
\vskip 0.2cm
{\Huge {\bf in the String Fusion Model Code}}\\
\vskip 0.8cm
E. G. Ferreiro and C. Pajares\\ {\sl Departamento de
F\'{\i}sica de  Part\'{\i}culas,\\
Universidade de Santiago de
Compostela,\\ 15706--Santiago de Compostela, Spain}

\vskip 2.0truecm
{\Large {\bf Abstract}}
\end{center}

String Fusion Model Code results for Pb--Pb central collisions at SPS energies
 including
the string fusion interaction mechanism are compared to the last
experimental data. Predictions for RHIC energies are also presented.

On the other hand, the evolution of the strangeness enhancement
ratio $E_s=\frac{<\Lambda>+4<K_s^0>}{3<\pi^->}$ with the energy and with the
atomic number of participant nuclei is discussed and related to the effects of
string fusion and high--coloured strings.

\vspace{4.5cm}
\begin{flushleft}
\end{flushleft}

\end{titlepage}
\newpage

\noindent{\bf 1. Introduction:}                                               
Strangeness enhancement has been proposed as a possible signal of the
transition towards a deconfined phase, the so--called Quark Gluon
Plasma (QGP) (\cite{Mu86,Raf91}). Using the String Fusion Model Code (SFMC,
(\cite{SFMC})) we have studied this effect for S--S and S--Ag central
collisions at SPS energies (\cite{Strange}).

In this paper we present the results obtained for Pb--Pb central collisions at
$p_{lab}=158$ GeV/c per nucleon
and we compare them to the last experimental data.

We also study the evolution of the 
strangeness enhancement
ratio $E_s=\frac{<\Lambda>+4<K_s^0>}{3<\pi^->}$ (\cite{Gaz95})
with the energy and with the
atomic number, and we 
propose other possible ratio to measure the strangeness,
$E^{\prime}_s=\frac{<\overline{\Lambda}>+4<K_s^0>}{3<\pi^->}$.

\vspace{0.5cm}
\noindent{\bf 2. The String Fusion Model Code:}
Several string models like the Dual Parton Model (DPM, (\cite{Cap94})) or
equivalently the Quark Gluon String Model (QGSM, (\cite{Kai84})) and models
based
 on
them like VENUS (\cite{Wer87}) have been very successful in describing particle
production in hadron--hadron, hadron--nucleus and nucleus--nucleus
interactions.
Monte Carlo versions of these models (\cite{SFMC,Wer87,Moh91,Ame93}) are in
reasonable agreement with most of the properties of soft multiparticle
production.

However, the enhanced generation of strange
particles (\cite{Roh93,And93,Aba91,GazQM93,Me93,Ger94}), in particular of
strange antibaryons and
$\phi$--mesons, that has been observed in the experiments at the CERN--SPS is
not explained in the models quoted above.

These models have introduced different mechanism in order to reproduce that
fenomenum.
In this way, in the DPM the creation of sea diquark--antidiquarks
pairs from the vacuum has been introduced, in addition to the usual
quark--antiquark pairs (\cite{Ran94}), and in VENUS the fusion of particles and
resonances into large
clusters which decay isotropically (\cite{Wer93}) has been considered.

The String Fusion Model Code (SFMC, (\cite{SFMC})) 
is a Monte Carlo code based on the 
QGSM that incorporates the possibility of string fusion.
In this code, at high energy a hadron or nucleus collision is assumed to be an
interaction between two clouds of partons, one from the projectile and the
other from the target. Each parton--parton inelastic interaction leads to the
creation of two colour strings. The strings have to be formed in pairs, with
oppositely coloured fluxes, because the projectile and the target should
remain colourless. We consider that the strings fuse when their transverse
positions come within a certain interaction area, of the same order as
parton--parton one. For present calculations it has been taken equal to 7.5 mb,
and only fusion of strings in groups of 2 has been included.
It is formally described by allowing partons to interact several
times, the number of interactions being the same for projectile and target. The
quantum numbers of the fused string are determined by the interacting partons
and
its energy--momentum is the sum of the energy--momenta of the ancestor
strings.
 The
colour charges of the fusing string ends sum into the  colour charge  of the
resulting string ends according to the $SU(3)$ composition laws.

The breaking of the fused string is due to the production of two (anti)quark
complexes with the same colour charges $Q$ and $\overline{Q}$ as those at the
ends
of the string.
The probability rate is given by the Schwinger formula
(\cite{Sch51,Car79,Gyu85}):
\beq
W \sim\\ K^2_{[N]}\\ exp(-\pi\\ M^2_t/K_{[N]})
\label{eq1}\ \ ,
\eeq
where $K_{[N]}$ is the string tension for the
$[N]$
$SU(3)$ representation proportional to the corresponding quadratic Casimir
operator $C^2_{[N]}$. In our case
\beq
C^2_{[3]}=4/3,\ \ C^2_{[6]}=10/3,\ \ C^2_{[8]}=3\ \ .
\eeq

Therefore, the [8] and [6] fused strings have a higher string tension,
giving rise to a larger heavy flavour production, in particular strangeness
production.

The string fusion mechanism also produces a decrease of the mean number of
strings. This leads to an overall reduction of multiplicities, specially of
mesons.

On the other hand, the fused string decays into more diquarks and antidiquarks
and into heavier quarks, so there will be an enhancement of strange particles,
specifically of strange baryons and antibaryons.

\vspace{0.5cm}
\noindent{\bf 3. Results:}
We have run our code in order to reproduce Pb--Pb central (impact parameter 
$b \leq 2.8$ fm) 
collisions at $p_{lab}=158$ GeV/c per nucleon.

Our results of negative charged particles, $K^0_s$ and protons for the string
fusion case compared to experimental data (\cite{NA49,NA44})
are shown in Figure 1.

It is seen that our results for $h^-$ are slightly higher than the experimental
ones in the central region. For proton production we obtain a good agreement,
specially when it is compared to other models
predictions (\cite{NA44}).

The experimental data on $\Lambda$ and $\overline{\Lambda}$ are not corrected
for particles originating from $\Xi$ and $\overline{\Xi}$ decays, that in our
code are taken into account separately.
The possible 
correction due to these decays is close to $30 \%$ (\cite{Alb94}). 
So in order to compare the data on
$\Lambda$ and $\overline{\Lambda}$ 
with our Monte Carlo results is necessary to
take the first ones reduced in this proportion.

As we can see in Ref. \cite{NA49}, the mean number of $\Lambda$ is around 10
times the one obtained for S--S central collisions, 
$<\Lambda>_{{\rm Pb-Pb}}= 10.5 \pm 3.7 <\Lambda>_{{\rm S-S}}$.
If we take the experimental value  $<\Lambda>_{{\rm S-S}}=9.4 \pm 1.0$ and we
correct it from $\Xi$ and $\overline{\Xi}$ contributions we obtain the
approximate value $<\Lambda>_{{\rm
Pb-Pb}} \sim 70$ from experimental data. 

The experimental ratio $\overline{\Lambda}/\Lambda$
for Pb--Pb central collisions is  $0.19 \pm 0.01$. We can calculate from this
relation a mean multiplicity for $\overline{\Lambda}$ around 13, if $\Xi$ and
$\overline{\Xi}$ contributions are not included.

Relative to the SFMC results (string fusion included), we have obtained 
the values   $<\Lambda>=37.4$ and $<\overline{\Lambda}>=10.5$.

It is observed that our result for $\Lambda$ is far away from experimental
data, due to the fact that we don't have rescattering or cascading mechanism
in the code. One of the main effects of cascading will be an enhancement of
$\Lambda$ production due to the processes:

\noindent
$\pi^-$p($\pi^0$n)$\rightarrow$$K^+$$\Sigma^-$, $\overline{K^0}$$\Lambda$ 
\ \ ;
\ \  $\pi^+$n($\pi^0$p)$\rightarrow$$K^+$$\Sigma^0$, $K^+$$\Lambda$,
                               $\overline{K^0}$$\Sigma^+$ \ \  and \ \ 
$K^-$p$\rightarrow$$\pi^0$$\Lambda$, $\pi^+$$\Sigma^-$ \ \  .

For $\overline{\Lambda}$ the code result when the string fusion mechanism is
included is close to the experimental one. Also, in the string fusion case we
obtained the relation $<\overline{\Lambda}>_{{\rm Pb-Pb}}=10.5
<\overline{\Lambda}>_{{\rm S-S}}$, while in the no fusion case this would be
$<\overline{\Lambda}>_{{\rm Pb-Pb}}=5.5
<\overline{\Lambda}>_{{\rm S-S}}$.

\vspace{0.5cm}
\noindent{\bf 4. Strangeness enhancement:}
The ratio 
\beq
E_s=\frac{<\Lambda>+4<K_s^0>}{3<\pi^->}
\eeq
 has been proposed 
in order to
study the total production of strangeness in nucleus--nucleus collisions
in a model independent way (\cite{Gaz95}). For an isospin zero system this
ratio is equivalent to the ratio
\beq
\frac{<\Lambda>+<K+\overline{K}>}{<\pi^->} \ \ .
\eeq

Our results for proton--proton
 and nucleus--nucleus
 collisions are shown in Figure 2 compared to
experimental data. We can observed that for proton--proton
 collisions at SPS
energies we obtain a value for $E_s$ much higher than the experimental
one. This is due to the fact that for $\Lambda$ and $\overline{\Lambda}$
production, our results for proton--proton
 collisions at SPS energies both in the
fusion and no fusion case are larger than the data because of the so--called
delayed threshold effect (\cite{Thres}).
At these energies the cross section for $\Lambda$ and $\overline{\Lambda}$
production is rising very sharply, so small changes in the energy will lead to
strong different results.

On the other hand, the mean number of strings created in a proton--proton
collision is very small, so the probability for them to fuse is low. Because
of this, the results don't change from the no fusion to the fusion case.
The string fusion effect grows with the energy and with the atomic number of
participant nuclei. So for nucleus--nucleus collisions the fusion
mechanism is more important and we find an enhancement of $E_s$, that is
larger for Pb--Pb collisions.

Other possibility is to study the ratio 
\beq
E^{\prime}_s=\frac{<\overline{\Lambda}>+4<K_s^0>}{3<\pi^->} \ \ .
\eeq
The obtained results are presented in Figure 3. For Pb--Pb central collisions
the experimental result coincides with the string fusion one.

The ratios  $\overline{\Lambda}/\pi^-$ and  $K^0_s/\pi^-$ are presented in
Tables 1 and 2 for the fusion and no fusion case compared to experimental
data. It is observed an enhancement of
 the ratio $\overline{\Lambda}/\pi^-$
from proton--proton to nucleus--nucleus collisions, that increases with the
atomic number of participant nuclei.
Nevertheless, for the experimental ratio
 $K^0_s/\pi^-$ there is only an
enhancement from proton-proton
 to nucleus--nucleus collisions, but it doesn't grow with the atomic
number.

It seems that there is an increase of strange antibaryon production but not
for strange meson one, which coincides with the code predictions. String
fusion mechanism has two antagonistic 
effects on strange production: on one hand
we have fused strings that decay into more strange particles, but on the
other hand these strings decay mainly into baryons and antibaryons, so the
overall dumping of multiplicities obtained as a consequence of the 
string fusion
affects mostly to meson production.

\vspace{0.5cm}
\noindent{\bf Discussion and conclusions:} 
It is important to take into account that in the SFMC only fusion of strings
in pairs has been included. 

Nevertheless the string fusion will be increased with the energy and the
atomic number, because the density of strings grows, as can be seen in Table
3.
So for Pb--Pb central collisions at SPS energies or for nucleus--nucleus
collisions at RHIC energies it is necessary to consider the possibility to
fuse the strings in bigger groups.

Even more, if the density of strings exceeds a critical value that can be
calculate knowing the radius of each string (0.2 fm), percolation of strings
becomes possible (\cite{Perco}).
The critical density necessary to have percolation is
$n_c=9$ strings/fm$^2$. Above it paths of overlapping strings (circles in the
transverse space, Figure 4) are formed through the whole collision area. Along
these paths the medium behaves as a colour conductor. 

The region where several strings fuse can be considered as a droplet of a
non--thermalized QGP. Percolation means that these droplets overlap and QGP
domain becomes comparable to nuclear size. This maybe a possible criterion
for the existence of a deconfining phase.

It is important to take this possibility into account when studying the
strangeness enhancement. In case of percolation the increase of strange
particle production would compete with a strong reduction of multiplicities
that would appear when many strings fuse into one and decay afterwards as an
only string.

In conclusion we want to thank A. Panagiotou for organizing such a nice
meeting and A. Capella and M. Gadzicki for useful comments and discussions.

\vspace{1cm}

\newpage
\noindent{\Large {\bf Table Captions}}

\vskip 0.5cm

\noindent{\bf Table 1.} Comparation between experimental data 
(\cite{NA49,NA44,Alb94}) 
and SFMC results
of the ratio $\overline{\Lambda}/\pi^-$ for p--p, S--S, S--Ag and
Pb--Pb central
collisions at SPS ($p_{lab}=200$ AGeV/c for p--p, S--S, S--Ag collisions and 
$p_{lab}=158$ AGeV/c for Pb--Pb collisions) and RHIC ($\sqrt{s}=200$ AGeV)
energies.

\noindent{\bf Table 2.} Comparation between experimental data 
(\cite{NA49,NA44,Alb94})
and SFMC results
of the ratio  $K^0_s/\pi^-$ for p--p, S--S, S--Ag and
Pb--Pb central  
collisions at SPS ($p_{lab}=200$ AGeV/c for p--p, S--S, S--Ag collisions and
$p_{lab}=158$ AGeV/c for Pb--Pb collisions) and RHIC ($\sqrt{s}=200$ AGeV)
energies.

\noindent{\bf Table 3.} Number of strings (upper numbers) and their densities
(fm$^{-2}$) (lower
numbers) in central p--p, S--S, S--U and Pb--Pb 
central collisions at SPS, RHIC and LHC
energies.

\newpage
\noindent{\Large {\bf Figure Captions}}

\noindent{\bf Figure 1.} Rapidity distributions of $h^-$ (upper left figure),
$K^0_s$ (upper
right figure), $p$ (lower left figure) and transverse mass distribution of
$h^-$
obtained with the SFMC code with string fusion and compared to experimental
data (black squares) (\cite{NA49,NA44}) 
for Pb--Pb central collisions at $p_{lab}=158$
AGeV/c.

\noindent{\bf Figure 2.} Ratio $E_s=\frac{<\Lambda>+4<K_s^0>}{3<\pi^->}$
vs. the atomic number of the target nucleus
 for p--p, S--S,
S--Ag and Pb--Pb central collisions at SPS (upper figure) and RHIC energies
(lower figure). The black stars correspond to SFMC results in the fusion case,
the empty stars correspond to SFMC results in the no fusion case and the 
points with error bars 
are the experimental data (\cite{NA49,NA44,Alb94}). For p--p collisions
the same result is obtained for the no fusion and the fusion case.

\noindent{\bf Figure 3.} Ratio
$E^{\prime}_s=\frac{<\overline{\Lambda}>+4<K_s^0>}{3<\pi^->}$
vs. the atomic number of the target nucleus
for p--p, S--S,
S--Ag and Pb--Pb central collisions at SPS (upper figure) and RHIC energies
(lower figure). The black stars correspond to SFMC results in the fusion case,
the empty stars correspond to SFMC results in the no fusion case and the 
points with error bars 
are the experimental data (\cite{NA49,NA44,Alb94}). 
                                                                         
\noindent{\bf Figure 4.} Strings in the transverse space. Each circle
corresponds to one string.

\vskip 0.5cm

\newpage
\begin{center}
{\bf Table 1}
\end{center}
\vskip 1cm
\begin{table}[hbtp]
\begin{center}
\begin{tabular}{cccc} \hline
Collision & Experiment & Without fusion & With fusion\\ \hline 
p--p at SPS & 0.005 & 0.0088 & 0.012 \\ 
S--S at SPS & 0.016 & 0.0055 & 0.013 \\ 
S--Ag at SPS & 0.016 & 0.005 & 0.013 \\ 
Pb--Pb at SPS & 0.020 & 0.004 & 0.019 \\ 
\hline
p--p at RHIC & & 0.0205 & 0.025 \\ 
S--S at RHIC & & 0.0185 & 0.028 \\ 
S--Ag at RHIC & & 0.0181 & 0.0285 \\ 
Pb--Pb at RHIC & & 0.0178 & 0.032  \\ 
\hline
\end{tabular}
\end{center}
\end{table}

\newpage

\begin{center}
{\bf Table 2}
\end{center}
\vskip 1cm
\begin{table}[hbtp]
\begin{center}
\begin{tabular}{cccc} \hline
Collision & Experiment & Without fusion & With fusion\\ \hline 
p--p at SPS & 0.065 & 0.091 & 0.092 \\ 
S--S at SPS & 0.11 & 0.091 & 0.101 \\ 
S--Ag at SPS & 0.09 & 0.092 & 0.096 \\ 
Pb--Pb at SPS & 0.11 & 0.93 & 0.11 \\ 
\hline
p--p at RHIC & & 0.10 & 0.10 \\ 
S--S at RHIC & & 0.10 & 0.12 \\ 
S--Ag at RHIC & & 0.10 & 0.12 \\ 
Pb--Pb at RHIC & & 0.10 & 0.13 \\ 
\hline
\end{tabular}
\end{center}
\end{table}

\newpage

\begin{center}
{\bf Table 3}
\end{center}
\vskip 1cm
\begin{table}[hbtp]
\begin{center}
\begin{tabular}{ccccc} \hline
\multicolumn{1}{c}{$\sqrt s \ \ \rm (AGeV)$} &
\multicolumn{4}{c}{Collision} \\ \hline 
& p--p & S--S & S--U & Pb--Pb  \\ 
\hline
19.4 & 4.2 & 123 & 268 & 1145 \\
& 1.3 & 3.5 & 7.6 & 9.5 \\
\hline
200 & 7.2 & 215 & 382 & 1703 \\
& 1.6 & 6.1 & 10.9 & 14.4 \\
\hline
5500 & 13.1 & 380 & 645 & 3071 \\
& 2.0 & 10.9 & 18.3 & 25.6 \\
\hline
\end{tabular}
\end{center}
\end{table}

\newpage
\centerline{\bf Figure 1}

\begin{figure}[hbtp]
\begin{center}
\mbox{\epsfig{file=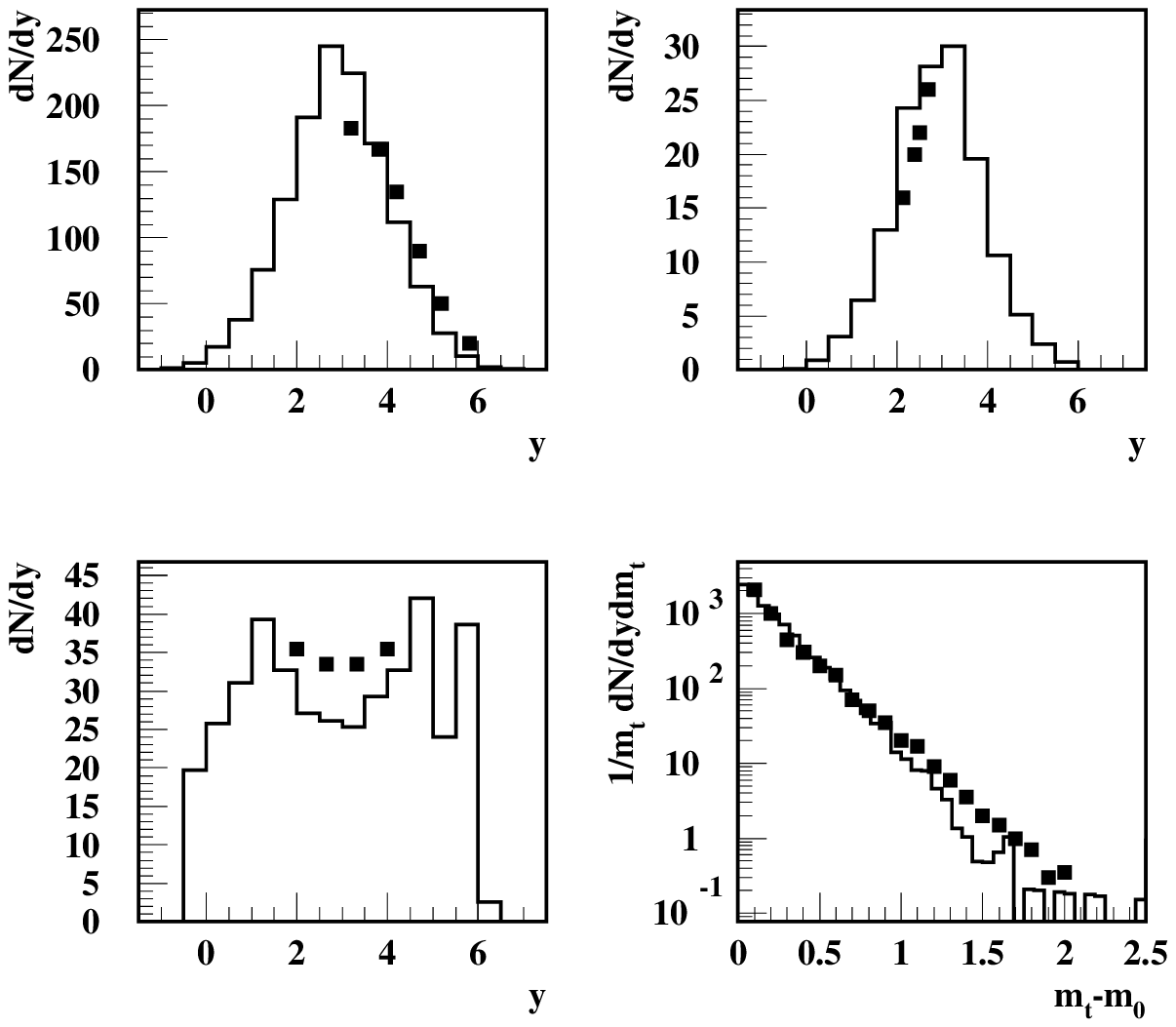,height=16cm}}
\end{center}
\end{figure}

\newpage
\centerline{\bf Figure 2}

\begin{figure}[hbtp]
\begin{center}
\mbox{\epsfig{file=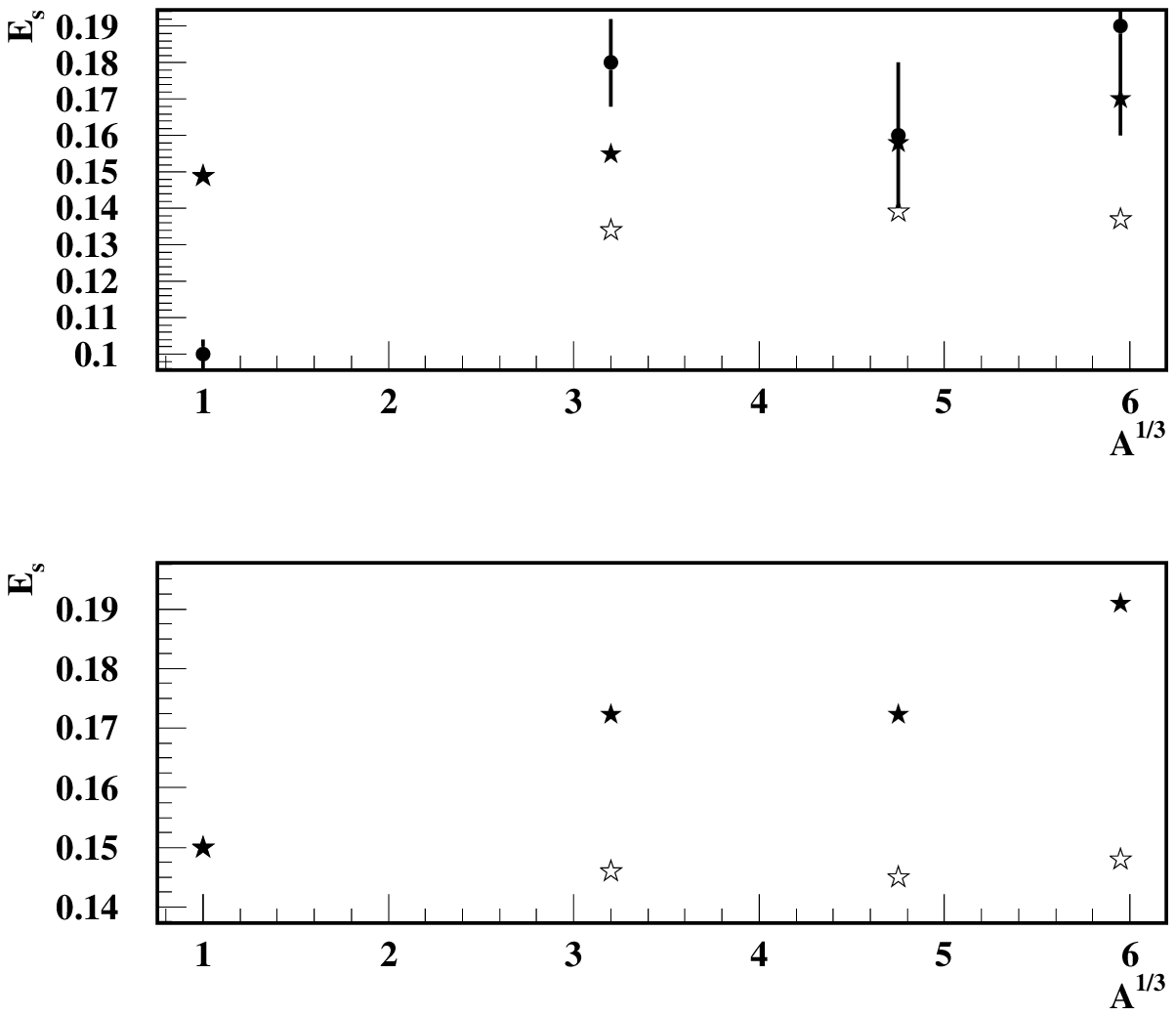,height=16cm}}
\end{center}
\end{figure}

\newpage
\centerline{\bf Figure 3}

\begin{figure}[hbtp]
\begin{center}
\mbox{\epsfig{file=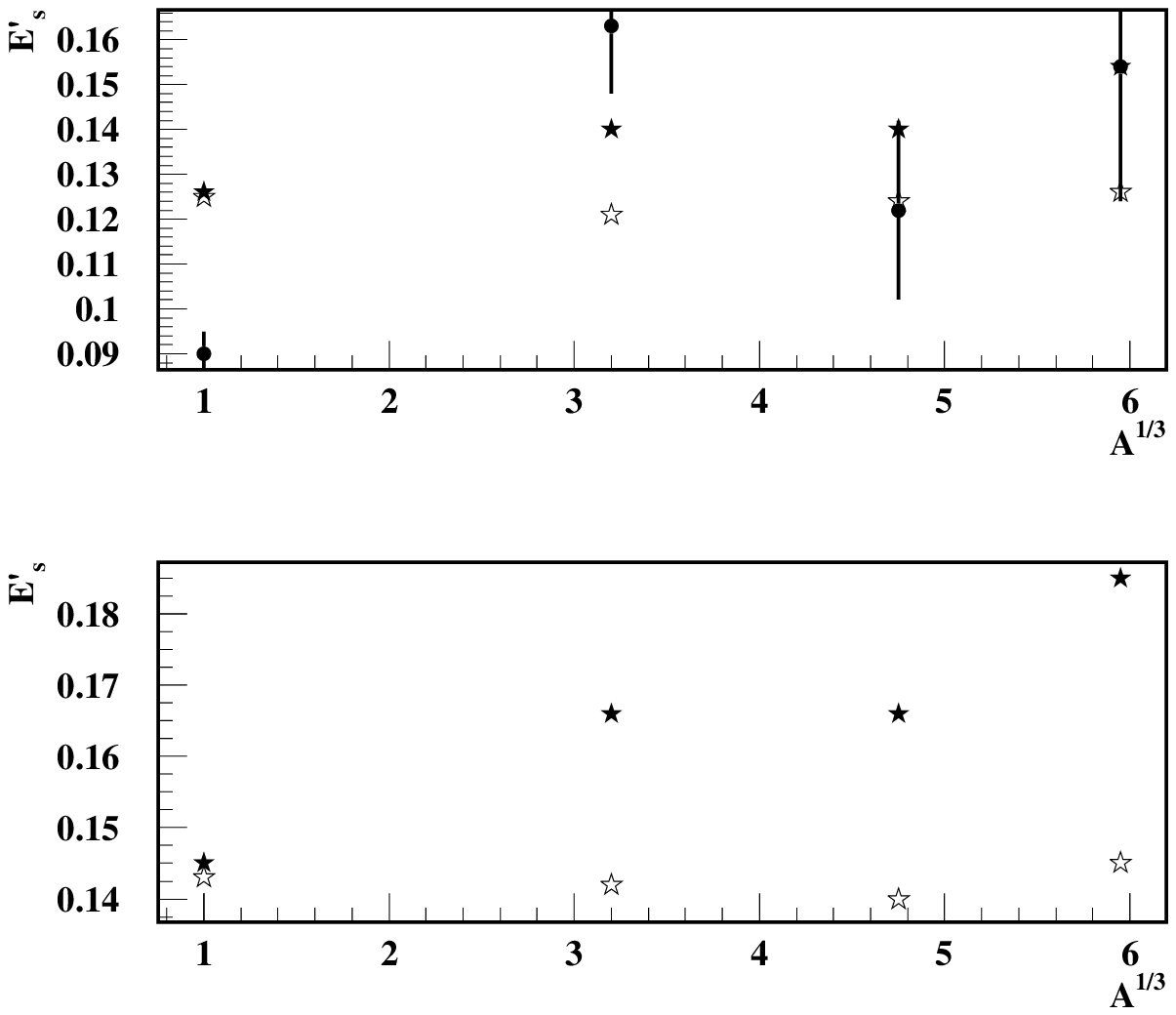,height=16cm}}
\end{center}
\end{figure}

\newpage
\centerline{\bf Figure 4}

\begin{figure}[hbtp]
\begin{center}
\mbox{\epsfig{file=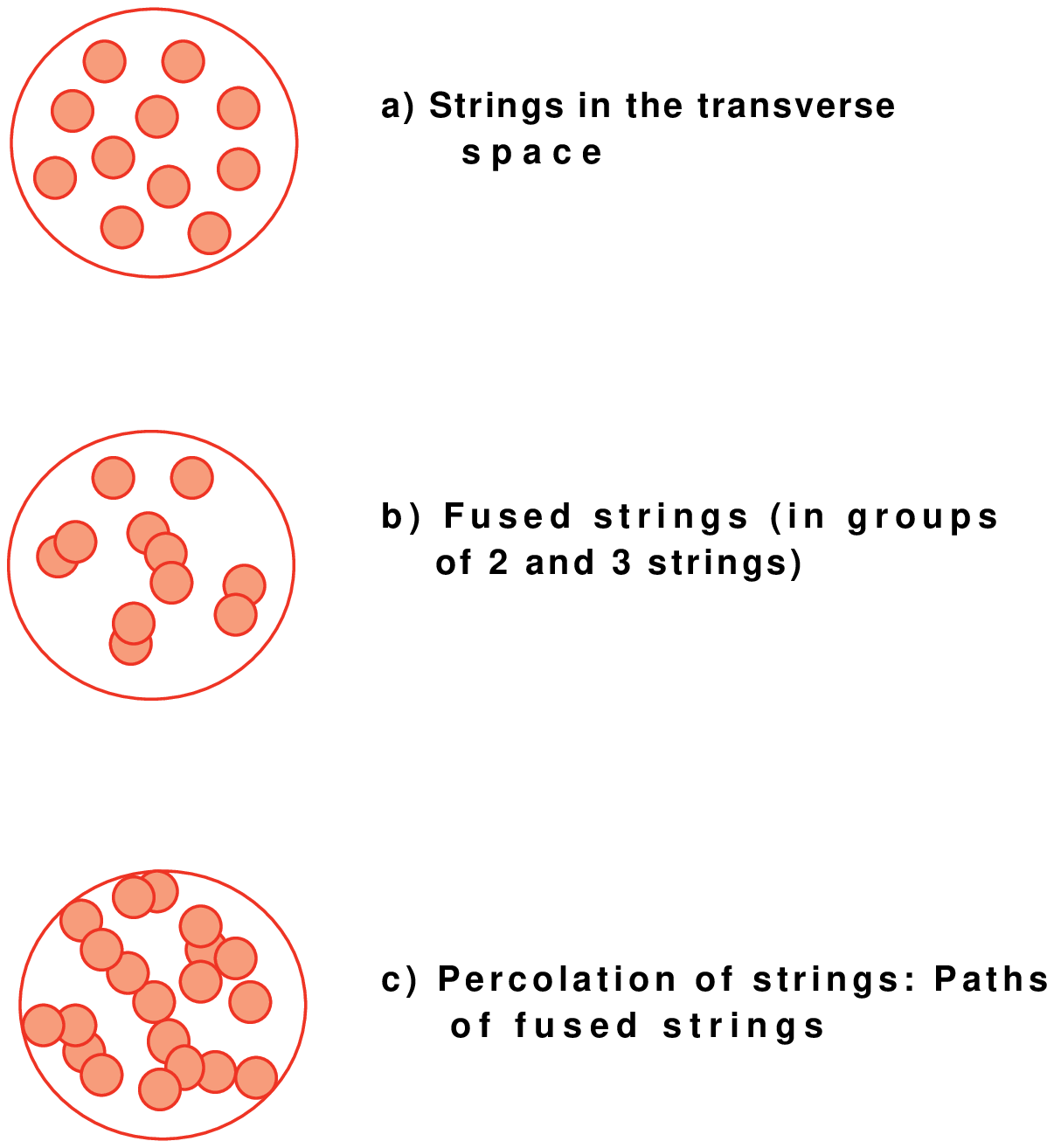,height=14cm}}
\end{center}
\end{figure}

\end{document}